\documentclass[numbers]{elsarticle}
\usepackage{latexsym}
\usepackage{natbib}
\usepackage[latin2]{inputenc}
\usepackage[hidelinks]{hyperref}
\usepackage{longtable, lineno}

\makeatletter
\def\ps@pprintTitle{%
	\let\@oddhead\@empty
	\let\@evenhead\@empty
	\def\@oddfoot{\centerline{\thepage}}%
	\let\@evenfoot\@oddfoot}
\makeatother

\begin{document}

\title{The {\tt braingraph.org} Database of High Resolution Structural Connectomes and the Brain Graph Tools}


\author[p,s]{Csaba Kerepesi\corref{cor2}}
\ead{kerepesi@pitgroup.org}
\author[p]{Balázs Szalkai\corref{cor2}}
\ead{szalkai@pitgroup.org}
\author[p]{Bálint Varga\corref{cor2}}
\ead{balorkany@pitgroup.org}
\author[p,u]{Vince Grolmusz\corref{cor1}}
\ead{grolmusz@pitgroup.org}
\cortext[cor1]{Corresponding author}
\cortext[cor2]{Joint first authors}
\address[p]{PIT Bioinformatics Group, Eötvös University, H-1117 Budapest, Hungary}
\address[u]{Uratim Ltd., H-1118 Budapest, Hungary}
\address[s]{Institute for Computer Science and Control, Hungarian Academy of Sciences, H-1111, Budapest, Hungary}

\date{}


\begin{abstract}
Based on the data of the NIH-funded Human Connectome Project, we have computed structural connectomes of 426 human subjects in five different resolutions of 83, 129, 234, 463 and 1015 nodes and several edge weights. The graphs are given in anatomically annotated GraphML format that facilitates better further processing and visualization. For 96 subjects, the anatomically classified sub-graphs can also be accessed, formed from the vertices corresponding to distinct lobes or even smaller regions of interests of the brain. For example, one can easily download and study the connectomes, restricted to the frontal lobes or just to the left precuneus of 96 subjects using the data. Partially directed connectomes of 423 subjects are also available for download. We also present a GitHub-deposited set of tools, called the Brain Graph Tools, for several processing tasks of the connectomes on the site \url{http://braingraph.org}. 
\end{abstract}

\maketitle

\section*{Introduction} 

Mapping all the inter-neuronal connections of the human brain with more than 80 billion neurons is not possible today. The discovery of connections between much larger areas of the gray matter of the human brain is feasible by applying diffusion tensor imaging (DTI) data acquisition and a subsequent data processing workflow. 

The NIH-funded large Human Connectome Project (HCP) \cite{McNab2013} regularly releases its high-quality functional- and diffusion MRI datasets of hundreds of healthy human subjects. One of the most interesting applications of the published data is the mapping of the connections of the human brain on a macroscopic level: State-of-the-art computational methods make possible of discovering neural fiber connections between 1015 anatomically identified gray matter areas (also called Regions of Interests, ROIs) of the brain \cite{Daducci2012,Fischl2012,Desikan2006,Tournier2012}. If we have DTI data from several human subjects, then these 1015 anatomically labeled cerebral areas can be corresponded between the individual cortices and sub-cortical gray matter areas through the subjects. 

We will get braingraphs or connectomes from this workflow if we identify the nodes (or vertices) with the 1015 ROIs, and we connect two such vertices by an edge if the workflow finds neural fibers, connecting the ROIs, corresponded to the two vertices. Therefore, by studying braingraphs we ignore the spatial orbits of the neural fibers in the white matter that connect the gray matter areas and can focus on the presence and the absence of connections between those ROIs. The edges of the graphs can be labeled by physical properties of the neural fibers connecting the corresponding ROIs.

Since the nodes of these graphs are corresponded to the very same set of 1015 anatomical areas, one can make comparisons between the braingraphs of individual subjects or groups of subjects in several ways and foci (e.g., \cite{Szalkai2015a,Szalkai2015,Kerepesi2015c,Szalkai2015c,Szalkai2016,Szalkai2016a,
Szalkai2016c,Kerepesi2015b,Kerepesi2016,Szalkai2016d}). 

Here we present the \url{http://braingraph.org} repository of connectomes, computed from the high-quality data of the Human Connectome Project \cite{McNab2013}, and some related software tools for the analysis and the visualization of braingraphs at the GitHub depository \url{https://github.com/}.

\section*{Discussion and Results}

The human braingraphs can be downloaded from the site \url{http://braingraph.org/download-pit-group-connectomes/}.

The following repositories are available:

\subsection*{Full set} The set contains the connectomes of 426 subjects from the Human Connectome Project's public data release \cite{McNab2013}. For each subject, we have prepared five graphs, with 83, 129, 234, 463 and 1015 nodes. Each graph is available as a separate GraphML file of name {\tt nnnnnn\_connectome\_scale\_xxx.graphml}. Here the first 6 digits refer to the subject ID from the Human Connectome Project's public release; and the last two or 3 digits to the vertex number in the graph. Scale 33 corresponds to 83 vertices, scale 60 to 129 vertices, scale 125 to 234 vertices, scale 251 to 463 vertices and scale 500 to 1015 vertices. 

In each file (i.e., in each graph) the following weights are given for each edge: 

{\tt FA\_std:} the standard deviation of the fractional anisotropies \cite{Basser2011} of the fiber(s); 

{\tt fiber\_length\_mean:} The mean of the fiber lengths, defining the graph edge, in millimeters.

{\tt fiber\_length\_std:} The standard deviation of the fiber lengths, defining the graph edge;

{\tt FA\_mean:} The mean of the fractional anisotropies \cite{Basser2011} of the fibers;

{\tt number\_of\_fibers:} the count of the fibers, defining the edge in question.

\subsection*{Directed graphs} The set contains 423 braingraphs with the 1015 nodes resolution. The edges of the graphs are directed by the Consensus Connectome Dynamics-based \cite{Kerepesi2015b,Kerepesi2016,Szalkai2016d} method, detailed in \cite{Szalkai2016d}. Every edge description field in GraphML format contains the directed status of the edge: if it is directed, then the source and target nodes are given, if the edge is not directed then it is noted as: {\tt  <edge directed="false" source=$u$ target=$v$> }, where $u$ and $v$ stand for vertex numbers, specified with anatomical information in the vertex-description field of the file. The edge weights are also noted as separate attributes.

\subsection*{Partial set} contains only the graphs of 96 subjects, otherwise the format is the same as the entries of the full set. This smaller set formed the basis for the studies \cite{Szalkai2015a} and \cite{Szalkai2015}.

\subsection*{Per-lobe connectomes} contain the subgraphs of the braingraphs of 96 subjects that are induced by the different lobes of the brain. That is, for each lobe, only those edges are listed that have both endpoints in the lobe. The edges carry the five weights, specified above.

\subsection*{Per-ROI connectomes} contain the subgraphs of the braingraphs of 96 subjects that are induced by the different ROIs of the brain. That is, for each ROI, only those edges are listed that have both endpoints in the ROI in question. All the edges carry the five weights, specified above. Small ROIs, even with just one vertex (e.g., the left amygdala) and large ROIs, with dozens of vertices (e.g., the right inferior-parietal lobule with 26 nodes) are also present in the set.

\subsection*{The Brain Graph Tools} is a GitHub-based repository of some software programs for the easy processing of the \url{http://braingraph.org}-deposited data. The depository can be accessed at \url{https://github.com/kerepesi/Brain-Graph-Tools}. There are three main set of tools on the site:

\begin{itemize}
\item The Budapest Reference Connectome workflow (with {\tt RefBrainGraph.pl}), contains the tools of preparing consensus connectomes from a set of braingraphs, as in \cite{Szalkai2016,Szalkai2015a}. Graphs, called $k$-consensus connectomes, contain the edges of $n$ connectomes that are present in $k$ or more braingraphs ($k\leq n$).

\item The Brain Diversity workflow contains the tools {\tt GenPreFile.pl} and {\tt BrainDiversity.pl}, and is capable of performing a related task: from $n$ connectomes, the individual variability of the edges of the distinct lobes or ROIs are calculated as in \cite{Kerepesi2015c}. The output contains interactive Google Charts visualizing the variabilities.

\item The Brain Evolution Workflow, containing {\tt GenPreFile.pl} and {\tt  BrainEvolution.pl}, is capable of comparing the random evolution of graph edges with the phenomenon, described as the ``Consensus Connectome Dynamics'' in \cite{Kerepesi2015b, Kerepesi2016}. The generated figures are also given as interactive Google Charts.

Further information is given in a README file at the site \url{https://github.com/kerepesi/Brain-Graph-Tools/blob/master/README}.

\end{itemize}

\begin{figure} [h!]
	\centering
	\includegraphics[width=5.2in]{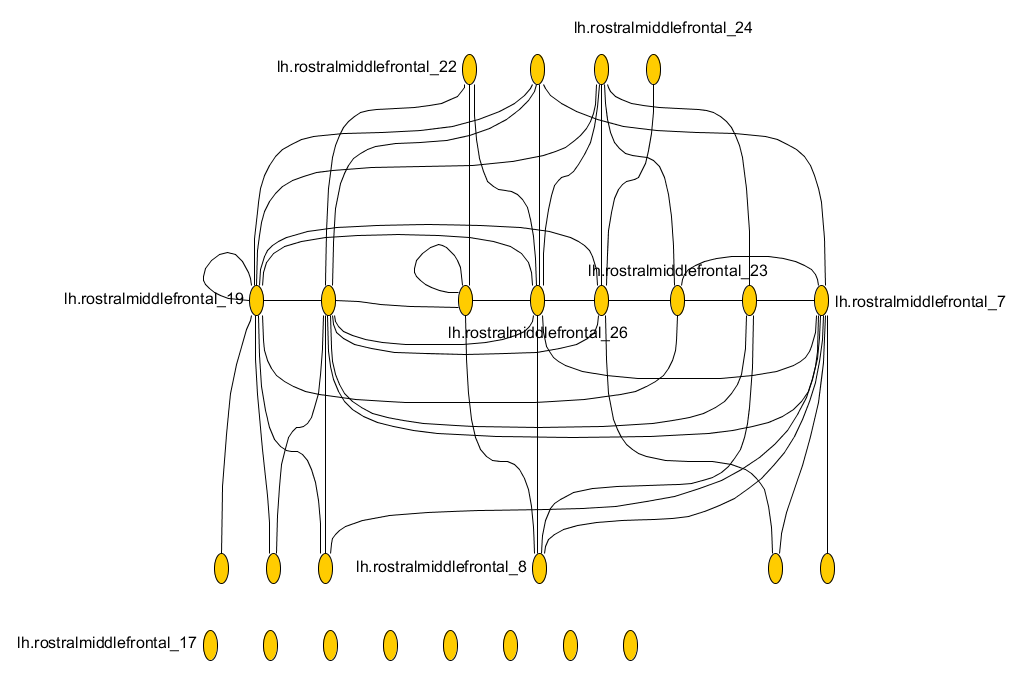}
	\caption{One example for the ROI subgraphs of the \url{http://braingraph.org/download-pit-group-connectomes/}, downloaded from the Repository {\bf Per-ROI connectomes}. The graph describes the connections within the left rostral-middle-frontal region-of-interest (ROI) of subject ID 127933, with the 1015-vertex resolution (counted for the whole graph), the original filename is {\tt lh.rostralmiddlefrontal-127933\_connectome\_scale500.graphml}. The graph contains 26 nodes and 44 edges, running between these nodes; some nodes are isolated. We show only some of the vertex-labels for clarity. The figure was prepared by the yED GraphML editor and viewer \url{https://www.yworks.com/products/yed}.}
\end{figure}

\section*{Methods}

The data source used was the Human Connectome Project's website: \url{http://www.humanconnectome.org/documentation/S500}  \cite{McNab2013}.

The connectomes were computed by using the Connectome Mapper Toolkit \cite{Daducci2012} \url{http://cmtk.org} for segmentation and partitioning. For tractography, we used the MRtrix processing tool \cite{Tournier2012} applying randomized seeding and the deterministic streamline method with a maximum of 20 000 fibers. 

The braingraphs are deposited in compressed form (by either 7-zip or zip) and are labeled by the HCP subject IDs. Some misconfigured systems contain 7-zip decompressing tools that do not decode properly the end-of-line characters; in this case, we suggest using a Linux system for decompressing the files.

\section*{Data availability:} The Human Connectome Project's MRI data is accessible at: 
\url{http://www.humanconnectome.org/documentation/S500} \cite{McNab2013}. 

\noindent The graphs (both undirected and directed) that were prepared by us from the HCP data can be downloaded at the site \url{http://braingraph.org/download-pit-group-connectomes/}. The Brain Graph Tools are available at \url{https://github.com/kerepesi/Brain-Graph-Tools}.

\section*{Acknowledgments}
Data were provided in part by the Human Connectome Project, WU-Minn Consortium (Principal Investigators: David Van Essen and Kamil Ugurbil; 1U54MH091657) funded by the 16 NIH Institutes and Centers that support the NIH Blueprint for Neuroscience Research; and by the McDonnell Center for Systems Neuroscience at Washington University. BS was supported through the new national excellence program of the Ministry of Human Capacities of Hungary.



\end{document}